\documentclass[12pt]{amsart}
\usepackage{amsthm,amssymb,latexsym}
\usepackage{graphicx}
\usepackage{fullpage}

\title{A Weak Definition of Delaunay Triangulation}
\author{Vin de Silva}
\thanks{This work has been supported by NSF grant DMS-0101364}
\address{Department of Mathematics, Building~380, Stanford University,
CA 94305-2125, USA. E-mail: {\tt silva@math.stanford.edu}}
\date{15 October, 2003}

\newcommand\Rr{\mathbb{R}}

\newcommand\Ss{S_+}
\newcommand\Hh{\mathcal{H}}

\newcommand\voro{\mathop{\rm Vor}}
\newcommand\dely{\mathop{\rm Del}}

\newcommand\commentout[1]{\marginpar{\tiny $\backslash$commentout}}

\theoremstyle{plain}
\newtheorem{Lemma}[equation]{Lemma}
\newtheorem{Theorem}[equation]{Theorem}
\newtheorem{Proposition}[equation]{Proposition}
\newtheorem{Corollary}[equation]{Corollary}

\theoremstyle{definition}
\newtheorem*{Example}{Example}
\newtheorem*{Definition}{Definition}

\theoremstyle{remark}
\newtheorem*{Remark}{Remark}

\begin{document}

\begin{abstract}
We show that the traditional criterion for a simplex to belong to the
Delaunay triangulation of a point set is equivalent to a criterion which
is \emph{a priori} weaker. The argument is quite general; as well as the
classical Euclidean case, it applies to hyperbolic and hemispherical
geometries and to Edelsbrunner's weighted Delaunay triangulation. In
spherical geometry, we establish a similar theorem under a genericity
condition. The weak definition finds natural application in the problem of
approximating a point-cloud data set with a simplical complex.
\end{abstract}

\maketitle
\section{Strong and weak witnesses}
\label{sec:introduction}

\begin{figure}[t]
\centerline{
\includegraphics[scale=0.85]{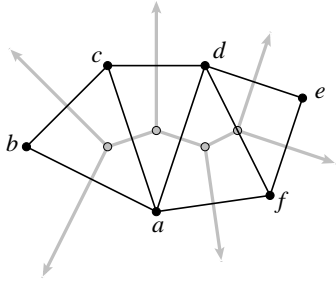}
}
\caption{The Voronoi diagram and Delaunay triangulation for a 6~point set.}
\label{fig:delaunay}
\end{figure}

Let $A \subset \Rr^n$ be a set of points, not necessarily finite. The
\emph{Voronoi diagram} of~$A$, denoted~$\voro(A)$, is the decomposition
of~$\Rr^n$ into Voronoi cells~$\{ V_a : a \in A\}$ defined as follows:
\[
V_a = \{ x \in \Rr^n: |x-a| \leq |x-b| \mbox{ for all } b\in A\}
\]
The dual of the Voronoi diagram is the \emph{Delaunay
triangulation}~$\dely(A)$, an abstract simplicial complex which contains
the $p$-simplex $[a_0 a_1 \ldots a_p]$ with vertices $a_0, a_1, \ldots,
a_p \in A$ if and only if $V_{a_0} \cap V_{a_1} \cap \cdots \cap V_{a_p}
\not= \emptyset$, for all $p \geq 0$.
When $A$ is a finite set of points in general position, there are no cells
with $p > n$, and $\dely(A)$ is geometrically realised as a triangulation
of the convex hull of~$A$. In this paper our interest is in the abstract
simplicial complex, which is defined under all circumstances. An example
of the geometrically-realisable case is given in
Figure~\ref{fig:delaunay}: the edges of the Delaunay triangulation of the
set $A =\{a,b,c,d,e,f\}$ are shown in black, and the Voronoi diagram is in
grey.

In order to state our theorem, we reformulate the definition of~$\dely(A)$
in terms of \emph{witnesses}.

\begin{Definition}
Let $\sigma = [a_0 a_1 \ldots a_p]$ be a $p$-simplex with vertices
in~$A$. We say that $x \in \Rr^n$ is a \emph{weak witness} for~$\sigma$
with respect to~$A$ if $|x - a| \leq |x - b|$ whenever $a \in \{ a_0, a_1,
\ldots, a_p\}$ and $b \in A \setminus \{a_0, a_1, \ldots, a_p\}$.
\end{Definition}
 
\begin{Definition}
Let $\sigma = [a_0 a_1 \ldots a_p]$ be a $p$-simplex with vertices
in~$A$. We say that $x \in \Rr^n$ is a \emph{strong witness} for~$\sigma$
with respect to~$A$ if $x$ is a weak witness for~$\sigma$ with respect
to~$A$ and additionally there is equality $|x - a_0| = |x - a_1| = \cdots
= |x - a_p|$.
\end{Definition}

\begin{figure}[t]
\centerline{
\includegraphics[scale=0.85]{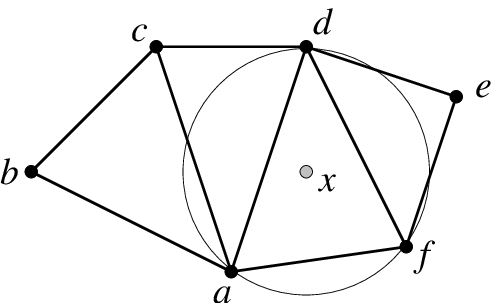}\hfill
\includegraphics[scale=0.85]{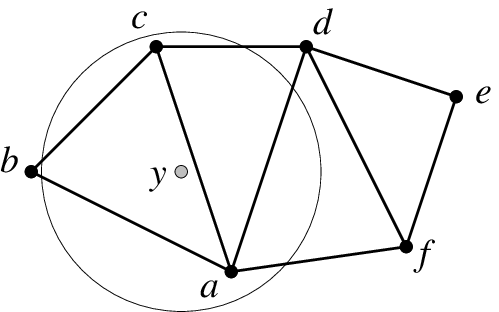}\hfill
\includegraphics[scale=0.85]{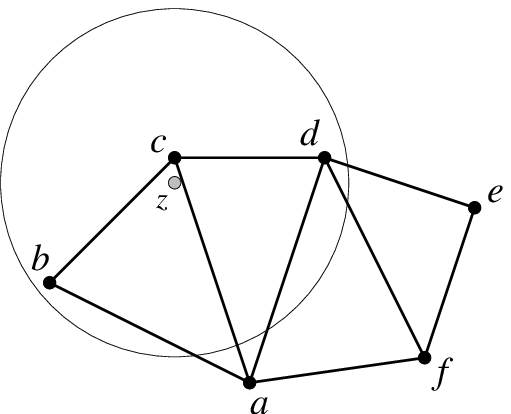}
}
\caption{Strong and weak witnesses.}
\label{fig:witnesses}
\end{figure}

Examples are given in Figure~\ref{fig:witnesses}.
In the left panel, $x$~is a strong witness for triangle~$[adf]$. In the
center panel, $y$~is a weak witness for edge~$[ac]$ and indeed for
triangle~$[abc]$. On the right, $z$~is a weak witness for~$[bcd]$. The
circumcenter of~$[bcd]$ lies close to~$a$, so $[bcd]$ does not have a
strong witness.

When the context is unambiguous, we may occasionally drop the phrase
``with respect to~$A$''. It follows from the definition of Voronoi cells
that $\sigma \in \dely(A)$ if and only if $\sigma$~has a strong witness.

\begin{Theorem}
\label{thm:main}
Let $A \subset \Rr^n$ and let $\sigma = [a_0 a_1 \ldots a_p]$ be a simplex
with vertices in~$A$. Then $\sigma$ has a strong witness if and only if
every subsimplex $\tau \leq \sigma$ has a weak witness.
\end{Theorem}

A consequence of this theorem is that the Delaunay triangulation can be
defined using weak witnesses only. Note that one direction is trivial: if
$x$~is a strong witness for~$\sigma$ then it is a strong (and hence weak)
witness for every subsimplex~$\tau\leq\sigma$.

\begin{Example}
If $abcd$ is a convex quadrilateral in the plane, then each of the
vertices $[a]$, $[b]$, $[c]$, $[d]$, each of the boundary edges $[ab]$,
$[bc]$, $[cd]$, $[da]$ and each of the triangles $[abc]$, $[bcd]$,
$[cda]$, $[dab]$ has a weak witness with respect to~$\{a,b,c,d\}$. If
$abcd$ is not cyclic, then exactly one of the diagonal edges $[ac]$,
$[bd]$ has a weak witness. 
For instance, in Figure~\ref{fig:witnesses}, edge~$[ac]$ has a weak witness
but edge~$[bd]$ does not.
The theorem predicts that the two triangles
containing that edge have strong witnesses; and indeed the Delaunay
triangulation consists of those two triangles joined along a common edge.
\end{Example}

The special case of an edge $\sigma = [a_0 a_1]$ is well-known, and
appeared in~\cite{Martinetz_Schulten_1994} to justify a construction of a
topology-approximating graph. When their construction is generalised to
topology-approximating simplicial
complexes~\cite{Carlsson_deSilva_2003sub}, the corresponding justification
is provided by the full theorem as it appears here. The terminology of
weak and strong witnesses was introduced in the latter paper.

\section{Proof of Theorem~\ref{thm:main}}
\label{sec:proof}

The proof of Theorem~\ref{thm:main} is an inductive generalisation of the
proof in~\cite{Martinetz_Schulten_1994} for the case of an edge. We
strengthen the statement of the theorem, in order to facilitate the
induction.

\begin{Theorem}
\label{thm:restated}
Let $A \subset \Rr^n$ and let $\sigma = [a_0 a_1 \ldots a_p]$ be a simplex
with vertices in~$A$. For every nonempty $I \subseteq \{a_0, a_1, \ldots,
a_p\}$, suppose there exists a weak witness~$x_I$ for the subsimplex
$\sigma_I \leq \sigma$ with vertex set~$I$. Then the convex hull of the
points~$x_I$ contains a strong witness for~$\sigma$.
\end{Theorem}

\begin{proof}
The proof proceeds by induction on~$p$. The base case $p=0$ is trivial.
Now suppose $p \geq 1$. Consider the subspace $H = \{ x \in \Rr^n: |x -
a_0| = |x - a_1| \}$ and the $(p-1)$-simplex $\tau = [a_1 a_2 \ldots
a_p]$. The following statements are easily seen to be equivalent:
\begin{eqnarray*}
&&
\mbox{$x \in H$ is a strong witness for~$\sigma$ with respect to~$A$}
\\
&\Leftrightarrow&
\mbox{$x \in H$ is a strong witness for~$\tau$ with respect to~$A$}
\\
&\Leftrightarrow&
\mbox{$x \in H$ is a strong witness for~$\tau$ with respect to~$A^\prime = A
\setminus \{a_0\}$}
\end{eqnarray*}
For every nonempty $J \subseteq \{a_1, \ldots, a_p\}$, we will find a weak
witness~$y_J \in H$ with respect to~$A'$ for the subsimplex~$\tau_J
\leq \tau$ having vertex set~$J$. By the inductive hypothesis, this
implies the existence of a strong witness~$x$ for~$\tau$ with respect
to~$A'$, in the convex hull of the points~$y_I$, and hence
in~$H$. The weak witnesses~$y_J$ are constructed as convex combinations of
the points~$x_I$, so $x$ itself belongs to the convex hull of the
points~$x_I$. This will complete the inductive step. It remains to locate
the weak witnesses~$y_J$.

The first case is $a_1 \in J$. Write $J = \{a_1\} \cup K$ where $K
\subseteq \{a_2, \ldots, a_p\}$, and consider weak witnesses $x_{0K}$,
$x_{1K}$ and $x_{01K}$ for the simplices with vertex sets $\{a_0\} \cup
K$, $\{a_1\} \cup K$ and $\{a_0,a_1\}\cup K$ respectively.
Suppose $|x_{01K} - a_1| \geq |x_{01K} - a_0|$. Since $|x_{1K} - a_1| \leq
|x_{1K} - a_0|$ it follows by continuity that there is some point~$y$ in
the closed line segment $\ell = [x_{01K}, x_{1K}]$ at which $|y - a_1| =
|y - a_0|$. Moreover, for every $a \in J$ and $b \in A' \setminus J$, the
convex inequality $|x - a| \leq |x - b|$ is valid at the endpoints
of~$\ell$ and hence throughout~$\ell$. It follows that $y_J = y$ is the
required weak witness for~$\tau_J$.
If instead $|x_{01K} - a_1| \leq |x_{01K} - a_0|$, there is a similar
argument with the roles of $a_0$ and~$a_1$ interchanged, giving
$y_J$~in~$[x_{01K}, x_{0K}] \cap H$. The two possibilities are illustrated
in Figure~\ref{fig:findingyJ}.
\begin{figure}[t]
\centerline{\hfill
\includegraphics{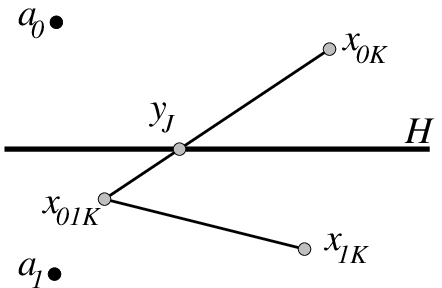}\hfill
\includegraphics{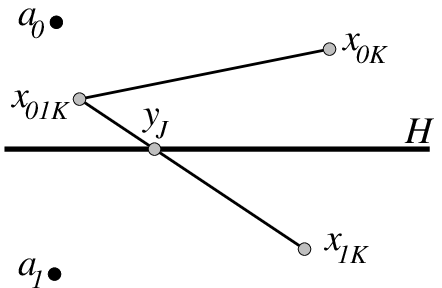}\hfill
}
\caption{Finding~$y_J$: the two possibilities in the case~$a_1 \in J$.}
\label{fig:findingyJ}
\end{figure}

The second case is $a_1 \not\in J$, so $\emptyset \not= J \subseteq \{a_2,
\ldots, a_p\}$. Consider weak witnesses $x_J$, $x_{0J}$ and~$x_{1J}$ for 
the simplices with vertex sets $J$, $\{a_0\} \cup J$ and $\{a_1\} \cup J$
respectively. Suppose $|x_J - a_0| \geq |x_J - a_1|$. Since $|x_{0J} -
a_0| \leq |x_{0J} - a_1|$, there is some point~$y$ in $\ell = [x_J,
x_{0J}]$ at which $|y - a_0| = |y - a_1|$. Moreover, for every $a \in J$
and $b \in A' \setminus J$, the inequality $|x - a| \leq |x - b|$ is valid
throughout~$\ell$. Thus the required weak witness is $y_J = y$. If instead
$|x_J - a_0| \leq |x_J - a_1|$, we can interchange the roles of $a_0$
and~$a_1$ to find $y_J$ in~$[x_J,x_{1J}] \cap H$.

This completes the proof of Theorem~\ref{thm:restated} and hence of
Theorem~\ref{thm:main}.
\end{proof}

\section{Variations}
\label{sec:variations}

The proof of Theorem~\ref{thm:restated} is quite general. Here is a set of
sufficient requirements, in the form of axioms.
\begin{enumerate}
\item
Let $R$ be a set with a notion of convexity; so every unordered pair $x,y
\in R$ determines a subset $[x,y] \subset R$ of intermediate elements.

\item
Let $A$ be a set. For every ordered pair $(a,b)$ in~$A$, there is a
\emph{convex} subset $R_{ab} \subset R$, interpreted as ``the set of
points whose distance from~$a$ is no greater than their distance
from~$b$''. Convex here means closed under taking intermediate elements.

\item
Define $H_{ab} = R_{ab} \cap R_{ba}$. If $x \in R_{ab}$ and $y \in R_{ba}$
then we require that $[x,y] \cap H_{ab} \not= \emptyset$. This rule stands
in for the intermediate value theorem.

\item
For all $a,b,c \in A$, we require that $H_{ab} \cap R_{ac} = H_{ab} \cap
R_{bc}$ and $H_{ab} \cap R_{ca} = H_{ab} \cap R_{cb}$. In other words, $a$
and~$b$ can be treated alike on the subspace of points equidistant from
them.

\end{enumerate}
One can define weak and strong witnesses in terms of the sets~$R_{ab}$,
and formulate analogous versions of Theorems \ref{thm:main}
and~\ref{thm:restated}. The proof goes through unchanged. Here are three
important special cases.
\begin{itemize}
\item
Let $R = \Hh^n$, hyperbolic $n$-space, and define $R_{ab} = \{ x \in \Hh^n
: d(x,a) \leq d(x,b)\}$ using the hyperbolic metric.

\item
Let $R = \Ss^n$, the open hemisphere in Euclidean $n+1$-space, defined by
\[
\Ss^n = \{ x \in \Rr^n :
x_1^2 + x_2^2 + \cdots + x_n^2 = 1, x_1 > 0 \}
\]
and set $R_{ab} = \{ x \in \Ss^n : d(x,a) \leq d(x,b)\}$ using the
geodesic metric.

\item
Let $R = \Rr^n$ and let $a \mapsto w_a$ be a real-valued function
on~$A$. Define $R_{ab} = \{ x \in \Rr^n : |x-a|^2 + w_a \leq |x-b|^2 +
w_b\}$.
\end{itemize}
In each case, $[x,y]$~is the set of points on the unique geodesic arc
from~$x$ to~$y$. The remaining axioms are easily verified; in particular
the sets~$R_{ab}$ are convex. It follows that weak witnesses can be used
to define the Delaunay triangulation in hyperbolic space and the Euclidean
hemisphere; as well as Edelsbrunner's weighted Delaunay
triangulations~\cite{Edelsbrunner_1995} in Euclidean space. In the latter
case, it is not necessarily true that there is a weak witness for every
0-dimensional simplex~$[a]$, so one must not fail to check this. In the
other cases, $a$~itself is always a witness for~$[a]$.

\begin{Remark}
A stronger version of Axiom~4 is \emph{transitivity}: $R_{ab} \cap R_{bc}
\subset R_{ac}$. This holds in all our examples.
\end{Remark}

\section{The Euclidean sphere}
\label{sec:sphere}

In this section and the next, we study the case of the Euclidean
sphere~$S^n$, with weak and strong witnesses defined in terms of the
intrinsic geodesic metric. The direct analogue of Theorem~\ref{thm:main}
is false, so we must weaken the claim. We begin by reformulating
Theorem~\ref{thm:main} itself.

\begin{Theorem}
\label{thm:disks}
Let $A \subset \Rr^n$ and let $\sigma = [a_0 a_1 \ldots a_p]$ be a simplex
with vertices in~$A$. Suppose that for every subsimplex~$\tau \leq \sigma$
there exists a closed Euclidean ball~$B_\tau$, such that $B_\tau$ contains
the vertices of~$\tau$ and the interior of~$B_\tau$ contains no other
points of~$A$. Then there exists a closed Euclidean ball~$B$, such that
the boundary of~$B$ contains the vertices of~$\sigma$ and the interior
of~$B$ is disjoint from~$A$.
\end{Theorem}

\begin{proof}
The existence of~$B_\tau$ is equivalent to the existence of a weak
witness~$x_\tau$ for~$\tau$: take $x_\tau$ to be the center of~$B_\tau$,
or conversely take $B_\tau$ to be the smallest ball centered on~$x_\tau$
which contains all the vertices of~$\tau$. Similarly, the existence of~$B$
is equivalent to the existence of a strong witness for~$\sigma$. In this
way Theorem~\ref{thm:disks} is equivalent to Theorem~\ref{thm:main}.
\end{proof}

For the sphere~$S^n$ we have the following result.

\begin{Theorem}
\label{thm:spheredisks}
Let $A \subset S^n$ and let $\sigma = [a_0 a_1 \ldots a_p]$ be a simplex
with vertices in~$A$. Suppose that for every subsimplex~$\tau \leq \sigma$
there exists a closed metric ball~$B_\tau$, such that $B_\tau$ contains
the vertices of~$\tau$ and the interior of~$B_\tau$ contains no other
points of~$A$. Suppose further that the union of the balls~$\{ B_\tau:
\tau \leq \sigma\}$ is not equal to the whole sphere~$S^n$. Then there
exists a closed metric ball~$B$, such that the boundary of~$B$ contains
the vertices of~$\sigma$ and the interior of~$B$ is disjoint from~$A$.
\end{Theorem}

\begin{proof}
Let $s \in S^n \setminus \bigcup\{B_\tau : \tau \leq \sigma\}$. Regard~$s$
as a vector in~$\Rr^{n+1}$ and let $V \subset \Rr^{n+1}$ be a hyperplane
perpendicular to~$s$. It is well known that the stereographic projection
$S^n \setminus \{s\} \to V$ maps closed metric balls in~$S^n$ which are
disjoint from~$s$ onto closed Euclidean balls in~$V$; and that this is a
bijective correspondence. Under this correspondence,
Theorem~\ref{thm:spheredisks} is equivalent to Theorem~\ref{thm:disks}.
\end{proof}

\begin{Example}
Theorem~\ref{thm:spheredisks} is false without the extra condition.
Consider the poles $n,s \in S^2$. Let $a_1, a_2, \ldots, a_p$ belong to a
circle of latitude~$C$, and let $a_0$ be any point lying north
of~$C$. Every simplex $\tau \leq \sigma = [a_0 a_1 \ldots a_p]$ then has a
weak witness: if $a_0 \in \tau$ then $n$~is a weak witness for~$\tau$, and
if $a_0 \not\in \tau$ then $s$ is a weak (and indeed a strong) witness
for~$\tau$. The corresponding disks are the northern and southern polar
caps bounded by~$C$; these cover~$S^2$ between them. If $p \geq 3$ then
$\sigma$ has no strong witness.
\end{Example}

The proof of Theorem~\ref{thm:restated} breaks down because of the failure
of convexity on~$S^n$. There is no unique shortest geodesic connecting a
pair of antipodal points on~$S^n$. More seriously the 0-sphere $S^0$ is
disconnected, so there is not even one geodesic on which to find the
intermediate point required by Axiom~3. The inductive argument works by
reducing the problem to subspheres of successively lower dimension, so the
$S^0$~case is unavoidable when $p > n$. On the other hand, for
subspheres~$S^k$ with $k \geq 1$, the failure of convexity is non-generic,
since it is only a problem when there is an antipodal pair of witnesses at
some stage in the proof. We will develop this idea in
Section~\ref{sec:sphere2}.

\begin{Corollary}
Suppose $A = \{ a_0, a_1, \ldots, a_{n+1}\} \subset S^n$ is not contained
in an $(n-1)$-sphere of any radius. For every $J \subset A$ suppose there
is a closed metric ball~$B_J$ containing~$J$ and whose interior is
disjoint from~$A\setminus J$. Then the balls~$B_J$ cover the sphere~$S^n$.
\qed
\end{Corollary}

\begin{Example}
For $0 < k < n$, identify $\Rr^n = \Rr^k \oplus \Rr^{n-k}$ with the
tangent space of~$S^n \subset \Rr^{n+1}$ at the north pole. For $1 \gg
\epsilon > \epsilon' > 0$, let $\{ a_0, a_1, \ldots, a_k\}$ be the
vertices of a regular $k$-simplex on the sphere of radius~$\epsilon$
in~$\Rr^k$, and let $\{ b_0, b_1, \ldots, b_{n-k}\}$ be the vertices of a
regular $(n-k)$-simplex on the sphere of radius~$\epsilon'$
in~$\Rr^{n-k}$, projected onto a neighbourhood of the north pole
in~$S^n$. If $\epsilon'$ is sufficiently close to~$\epsilon$, it can be
checked that every simplex $\tau \leq \sigma = [a_0 \ldots a_k b_0 \ldots
b_{n-k}]$ has a small witness disk~$B_\tau$ contained in a neighbourhood
of the north pole, \emph{except} for $\tau_a = [a_0 a_1 \ldots
a_k]$. Since~$\sigma$ does not have a strong witness, it follows that
every witness disk for the $k$-simplex~$\tau_a$ is necessarily large, and
in particular contains the entire southern hemisphere.
\end{Example}

\section{The Euclidean sphere, continued}
\label{sec:sphere2}

We will prove the following theorem.

\begin{Theorem}
\label{thm:spheregeneric}
For a generic finite subset $A \subset S^n$, and $p \leq n$, a simplex
$\sigma = [a_0 a_1 \ldots a_p]$ with vertices in~$A$ has a strong witness
if and only if every subsimplex $\tau \leq \sigma$ has a weak witness.
Here ``generic'' means that, for all $k < n$, no set of $k+3$ points
in~$A$ is contained in a $k$-sphere (of any radius).
\end{Theorem}

Thus if $A \subset S^n$~is generic we can use weak witnesses to define its
Delaunay triangulation in~$S^n$ correctly as far as the $n$-skeleton. On
the other hand, the true Delaunay triangulation is at most $n$-dimensional
if $A$~is generic: the existence of an $(n+1)$-cell would imply that its
$n+2$~vertices lay on a common $(n-1)$-sphere. The conclusion is that the
Delaunay triangulation in~$S^n$ of a generic set~$A$ is equal to the
$n$-skeleton of the complex defined using weak witnesses. Since a
randomly-chosen $A \subset S^n$ is generic with probability~1, this is a
satisfactory result.

\begin{Definition}
Let $A \subset S^n$ (not necessarily finite), and let $\sigma = [a_0 a_1
\ldots a_p]$ be a $p$-simplex with vertices in $A$. We say that $x \in
S^n$ is a \emph{robust witness} for~$\sigma$ with respect to~$A$ if
$x$~has a neighbourhood in which every point is a weak witness
for~$\sigma$ with respect to~$A$.
\end{Definition}

The next proposition may clarify the situation.

\begin{Proposition}
\label{prop:robustness}
For $n \geq 1$, let $A \subset S^n$ be a finite set and let $\sigma = [a_0
a_1 \ldots a_p]$ be a $p$-simplex with vertices in~$A$. Then $x$ is a
robust witness for~$\sigma$ if and only if there is strict inequality
$d(x,a) < d(x,b)$ whenever $a \in \{a_0, a_1, \ldots, a_p\}$ and $b \in A
\setminus \{a_0,a_1,\ldots,a_p\}$.
\end{Proposition}

\begin{proof}
Assume first that $x$ is a robust witness, and $a \in \{a_0, a_1, \ldots,
a_p\}$ and $b \in A \setminus \{a_0,a_1,\ldots,a_p\}$. Suppose $d(x,a) =
d(x,b) < \pi$. (The possibility $d(x,a) = d(x,b) = \pi$ is ruled out since
$x$~has only one antipode.) Let $y \not= x,b$ be any point on the minimal
geodesic~$[x,b]$. Then
\[
d(y,b) = d(x,b) - d(x,y)
\]
but
\[
d(y,a) < d(x,a) - d(x,y)
\]
since $y$ is not on the minimal geodesic~$[x,a]$. Thus $d(y,b) < d(y,a)$
for points~$y$ which can be arbitrarily close to~$x$. Since this would
contradict the robustness of~$x$ it must be that $d(x,a) < d(x,b)$ in all
cases.

In the converse direction, note that the inequalities $d(x,a) < d(x,b)$
are open conditions and there are finitely many of them. If all of the
appropriate inequalities are satisfied at~$x$ (making $x$ a weak witness),
then they are also satisfied on an open neighbourhood of~$x$. Thus $x$~is
a robust witness.
\end{proof}

\begin{Theorem}
\label{thm:robustwitness}
Let $A \subset S^n$ (not necessarily finite), let $p \leq n$, and let
$\sigma = [a_0 a_1 \ldots a_p]$ be a $p$-simplex with vertices in~$A$. If
every subsimplex $\tau \leq \sigma$ has a robust witness, then
$\sigma$~has a strong witness.
\end{Theorem}

\begin{proof}
For $k = 0, 1, \ldots, p$, let $\sigma_k = [a_k a_{k+1} \ldots a_p]$, let
$A_k = A \setminus \{ a_i: i < k\}$ and let $H_k$ be the great sphere
defined by the equations:
\[
d(x,a_0) = d(x,a_1) = \cdots = d(x,a_k)
\]
Note that $\dim(H_k) \geq n - k$. We will show inductively that every
subsimplex $\tau \leq \sigma_k$ has a robust witness in~$H_k$ with respect
to~$A_k$. The case $k=0$ is our hypothesis, and the case $k=p$ implies the
conclusion, since a witness~$x \in H_p$ (robust or otherwise) for~$[a_p]$
with respect to~$A_p$ is automatically a strong witness for~$\sigma$.

Suppose the assertion is proved for some $k < p$. For each $\tau \leq
\sigma_k$  we have a robust witness~$x_\tau \in H_k$ with respect
to~$A_k$. We may assume that no two of the $x_\tau$ are antipodal; since
$\dim(H_k) \geq 1$ we can perturb the points~$x_\tau$ to make this
true.
We now apply the argument in the proof of Theorem~\ref{thm:restated} to
find weak witnesses~$y_\tau \in H_{k+1}$ for $\tau \leq \sigma_{k+1}$ with
respect to~$A_{k+1}$. If an inequality $d(x,a) \leq d(x,b)$ is satisfied
on neighbourhoods of points $x_0, x_1$ which are not antipodal, then it is
satisfied on a neighbourhood of the minimal geodesic interval~$[x_0,x_1]$.
Thus the new witnesses $y_\tau$ are themselves robust. This completes the
inductive step.
\end{proof}

If $A$~is generic we can manufacture robust witnesses from weak witnesses.

\begin{Lemma}
\label{lemma:weak2robust}
Suppose $A \subset S^n$ is a finite set satisfying the condition that, for
all~$k < n$, no set of $k+3$~points in~$A$ is contained in a
$k$-sphere. Let $\sigma$ be a simplex with vertices in~$A$. If $\sigma$
has a weak witness, then $\sigma$ has a robust witness.
\end{Lemma}

\begin{proof}
Let $x$ be a weak witness for~$\sigma$. If $x$ is not robust, then there
are points $a_1, a_2, \ldots, a_k$ which are vertices of~$\sigma$, and
$b_1, b_2, \ldots, b_l \in A$ which are not, such that
\[
d(x,a_1) = \cdots = d(x,a_k) = d(x,b_1) = \cdots = d(x,b_l),
\]
so all these points lie on a $(n-1)$-sphere~$C$ centered on~$x$. The other
vertices of~$\sigma$ may be assumed to lie inside~$C$, and the remaining
points of~$A$ outside~$C$. By the condition, $k+l < n+2$.

Regarding $S^n$ as a subspace of~$\Rr^{n+1}$, consider the $n$-plane~$V$
defined by $V \cap S^n = C$. This can naturally be identified with the
tangent space of~$S^n$ at~$x$. We will construct a vector~$w \in V$, and
show that, for a small perturbation $x' = x + \epsilon w + O(\epsilon^2)$,
there are strict inequalities $d(x',a_i) < d(x',b_j)$ for all~$i,j$. When
$\epsilon$ is small enough, the relevant inequalities for the remaining
points of~$A$ remain unchanged, so $x'$~will be the required robust
witness.

Inside~$V$, we claim that the convex hull of $\{ a_1, \ldots, a_k, b_1,
\ldots, b_l\}$ is an embedded $(k+l-1)$-simplex. If this were false, then
the $k+l$~points would lie on a $(k+l-2)$-plane in~$\Rr^n$ and hence on a
$(k+l-3)$-sphere in~$S^n$, contradicting the hypothesis. It follows that
the convex hulls of~$\{a_1, a_2, \cdots, a_k\}$ and $\{b_1, b_2, \ldots,
b_l\}$ are disjoint, so they can be strictly separated by an
$(n-1)$-plane~$W$. Let $w$~be the unit normal vector of $W < V$, pointing
towards the half-space containing the points~$a_i$. See
Figure~\ref{fig:perturb}.
\begin{figure}
\hfill
\includegraphics{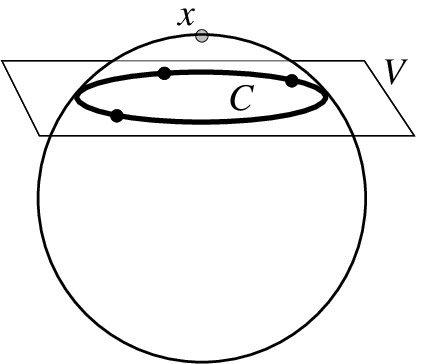}\hfill
\includegraphics{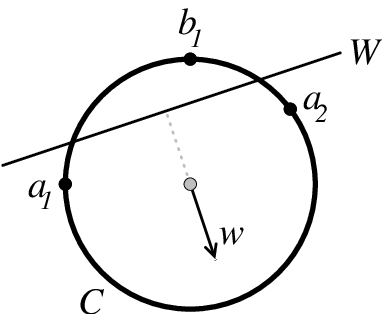}\hfill{}
\caption{[left] The $n$-plane~$V$ determined by the sphere~$C$; [right]
the separating hyperplane~$W$ determines a suitable direction for
perturbation.}
\label{fig:perturb}
\end{figure}

We now determine the effect of the perturbation $x' = x + \epsilon w +
O(\epsilon^2)$ along~$S^n$, on the distances. It is simplest to compute
the derivative of the squared Euclidean distance function $f_a: \epsilon
\mapsto |x + \epsilon w - a|^2$, namely $f_a'(0) = - 2\langle w, a
\rangle$ (since $\langle w, x \rangle = 0$). Since $w$
was chosen as the normal to a separating hyperplane, we deduce the
inequality $f_{a_i}'(0) - f_{b_j}'(0) = 2\langle w, b_j-a_i \rangle < 0$
for all~$i,j$. Thus for small~$\epsilon$ we have $|x'-a_i|^2 < |x'-b_j|^2$
and hence $d(x',a_i) < d(x',b_j)$ since there is a monotonic relationship
between the two distance functions. This completes the proof.
\end{proof}

\begin{proof}[Proof of Theorem~\ref{thm:spheregeneric}]
We have the following sequence of implications:
\begin{eqnarray*}
\mbox{$\sigma$ has a strong witness}
&\Rightarrow&
\mbox{every $\tau \leq \sigma$ has a weak witness}\\
\mbox{[by Lemma~\ref{lemma:weak2robust}, generically]} &\Rightarrow&
\mbox{every $\tau \leq \sigma$ has a robust witness}\\
\mbox{[by Theorem~\ref{thm:robustwitness}, since $p \leq n$]} &\Rightarrow&
\mbox{$\sigma$ has a strong witness}
\end{eqnarray*}
This proves the theorem.
\end{proof}

\section{Application}
\label{sec:application}

It is a standard problem in computational geometry to recover the topology
of a manifold $M \subset \Rr^n$ from a discrete sample of points $A
\subset M$. Recovery is often taken to mean the construction of a
simplical complex with vertices in~$A$, which is homeomorphic to the
unknown space~$M$. This problem has a rich literature.

When $M$ is regarded as known, it is natural to consider is the restricted
Delaunay triangulation $\dely(A,M)$, which is dual to the partitioning
of~$M$ by Voronoi cells. Unlike the full Delaunay complex~$\dely(A)$,
which is always contractible, $\dely(A,M)$ carries information about the
topology of~$M$ and may even be homeomorphic to it, if $A$ is sampled
sufficiently finely. According to the traditional definition, a
simplex~$\sigma$ with vertices in~$A$ belongs to~$\dely(A,M)$ if there is
a strong witness for~$\sigma$ on~$M$.

When $M$ is not known, one may attempt to approximate~$\dely(A,M)$. This
approach, introduced in~\cite{Martinetz_Schulten_1994} for approximation
by graphs, is extended to higher dimensional complexes
in~\cite{Carlsson_deSilva_2003sub}. The idea is to select a subsample~$L
\subset A$ to serve as the vertex set, and to let the remaining points
stand in for the unknown manifold~$M$. Unfortunately, it is useless to
define~$\dely(L,A)$ by mimicking the traditional definition
of~$\dely(L,M)$, since strong witnesses in~$A$ exist with
probablility~0. The solution is to replace definitions involving strong
witnesses with definitions involving weak witnesses, which do exist with
nonzero probability.

The goal of this paper has been to establish the credentials of weak
witness definitions, proving equivalence in the archetypal cases of
Delaunay triangulation in Euclidean, hyperbolic and spherical geometries,
and paying attention to some of the pitfalls. All this may be regarded as
groundwork for the following question: under what conditions is the
weak-witness complex~$\dely_w(L,A)$ homeomorphic (or homotopy equivalent)
to~$M$?

\bibliography{../BibTeX/vin}
\bibliographystyle{alpha}
\end{document}